\documentclass[
  journal=pasa
]{cup-journal}
\usepackage{graphicx, subfigure}
\usepackage{amsmath}
\usepackage{gensymb}
\usepackage[nopatch]{microtype}
\usepackage{booktabs}
\usepackage{hyperref}
\usepackage{orcidlink}

\newcommand\orcidicon[1]{\href{https://orcid.org/#1}{\mbox{\scalerel*{
\begin{tikzpicture}[yscale=-1,transform shape]
\pic{orcidlogo};
\end{tikzpicture}
}{|}}}}
\title{Joint RV-Astrometric Analysis of two-planet system HD~65216: Prospects for \textit{Gaia} DR4}

\author{Alexander~L.~Wallace\orcidlink{0000-0002-6591-5290}}
\affiliation{Centre for Astrophysics, University of Southern Queensland, C, Darling Heights, QLD 4350, Australia}
\email[Alexander~Wallace]{Alexander.Wallace@unisq.edu.au}
\author{Robert~A.~Wittenmyer\orcidlink{0000-0001-9957-9304}}
\affiliation{Centre for Astrophysics, University of Southern Queensland, C, Darling Heights, QLD 4350, Australia}
\author{Jonathan~Horner\orcidlink{0000-0002-1160-7970}}
\affiliation{Centre for Astrophysics, University of Southern Queensland, C, Darling Heights, QLD 4350, Australia}


\begin{document}

\begin{abstract}
We provide updated mass and orbital constraints on the planetary system around HD~65216, consisting of at least two giant planets on eccentric orbits.  We first analyse new RV data to constrain the periods and eccentricities of both planets, before applying the \textit{Hipparcos}-\textit{Gaia} proper motion anomaly to constrain the inclination of HD~65216~c.\\
From this joint analysis, we find that HD~65216~c is most likely in an eccentric orbit ($e$ = 0.42$\pm$0.05) with a longer period (7007$^{+192}_{-161}$\,days) than previously determined.  The proper motion anomaly is consistent with a bimodal distribution of inclinations, with peaks at sin$i\sim$0.5, producing true masses of 2.23$^{+0.96}_{-0.77}$\,M$_{\mathrm{J}}$ and 3.32$^{+1.40}_{-1.16}$\,M$_{\mathrm{J}}$ for HD~65216~b and c respectively, assuming the planets are co-planar.\\
Finally, we simulate \textit{Gaia} DR4 epoch astrometry and the DR3-DR4 proper motion anomaly.  We find that these methods will be sensitive enough to further constrain orbital parameters and true mass.
\end{abstract}

\section{Introduction}
%
The first planets discovered orbiting Sun-like stars \citep[e.g.][]{mayor1995jupiter,47Uma,70Vir,Butler3planet} were found using the radial velocity (RV) technique -- observing the line-of-sight wobble of a star to infer the presence of an unseen planet\footnote{For a detailed overview of the various methods used to discover exoplanets, we direct the interested reader to \citet{Handbook}, and references therein.}. In the three decades since, that technique has proven to be extremely powerful -- only outperformed in pure discovery numbers by the transit method, which has the advantage of being able to study thousands or tens of thousands of stars at once.\\
At the time of writing, the RV technique has yielded 1186 confirmed exoplanet discoveries\footnote{Value obtained from the NASA Exoplanet Archive \citep{ExoplanetArchive}, through a search by Discovery Method, on 26th June, 2026; at the time, the total number of confirmed exoplanet discoveries stood at 6298 planets.}. In addition, the technique has been widely employed in the characterisation of exoplanets discovered through other methods \citep[e.g.][]{Char1,Char2,Char3,Char4,Char5}, since it offers a direct mechanism by which a planet's mass can be measured. The more massive the planet, the larger the wobble it will induce in the motion of its host star. The RV method provides strong constraints on orbital periods, eccentricities, and companion minimum masses across a broad range of separations, and as a result remains an excellent tool for the characterisation of planetary systems.
%
%
However, the RV method only provides a minimum mass ($M\textrm{sin}i$), with the true mass and orbital inclination to the line of sight remaining unknown.  This can be solved by combining RV measurements with astrometry, which provides information on a star's reflex motion on the plane of the sky, motion that is perpendicular to that detected by the RV method \citep{brown15}. This information therefore complements the RV measurements, allowing the direct determination of the inclination of the orbits relative to the plane of the sky, and from this, the true mass of the planets.  The \textit{Gaia} space telescope \citep{perryman2014astrometric} has already proven the effectiveness of this technique in determining the true mass of high-mass planets and brown dwarfs \citep[e.g.][]{stevenson23,unger23,wallace24} on periods of $\sim$1--5\,years.\\
%
%
While RV sensitivity generally decreases with increasing orbital period, the astrometric reflex signal increases approximately linearly with orbital separation.  Long-period giant planets are therefore particularly well suited to astrometric characterisation. However, companions with orbital periods substantially exceeding the \textit{Gaia} observing baseline can remain difficult to detect from epoch astrometry alone, due to the fact that the observing baseline doesn't cover a full orbit.  The orbital motion is therefore difficult to distinguish from the star's proper motion \citet{lammers26}. In such systems, the reflex motion instead manifests as a discrepancy between short-term and long-term proper motions. The Hipparcos--Gaia Catalogue of Accelerations (HGCA) \citep{brandt21} provides a homogeneous framework for exploiting this effect over a baseline of more than two decades. In recent years, this has been the widest application of \textit{Gaia} astrometry to studies of exoplanets \citep[e.g.][]{damasso20,venner21,llop21,barbato25,pinamonti26}. \\
%
%
HD~65216 is a nearby G5V star \citep{houk78} hosting at least two giant planets identified through RV monitoring \citep{wittenmyer19}. The system also contains two M7--8 stellar companions 253\,AU away from the primary \citep{mugrauer07}, forming a triple system.  The study by \citet{wittenmyer19} determined the inner planet, HD~65216~b, has a minimum mass of 1.30$\pm$0.06\,M$_{\mathrm{J}}$ and a period of 578$\pm$1\,days, while the outer companion, HD~65216~c, has a minimum mass of 2.03$\pm$0.11\,M$_{\mathrm{J}}$ and a period of 5370$\pm$20\,days. Owing to its wide orbit and comparatively large minimum mass, HD~65216~c is an excellent candidate for astrometric detection through the HGCA proper motion anomaly. \\
In this paper, we combine HARPS RV observations with astrometric constraints from the HGCA to refine the orbital architecture of the HD~65216 system.  Section~\ref{sec:rv} presents an updated RV analysis with new orbital parameters for both planets.  In Section~\ref{sec:astro}, we model the HGCA proper motion anomaly to derive constraints on the inclination and true mass of HD~65216~c. Finally, in Section~\ref{sec:dr4}, we investigate the detectability of the system with \textit{Gaia} DR4 epoch astrometry and the expected DR3--DR4 internal proper motion anomaly.
\section{Radial Velocity Analysis}
\label{sec:rv}
A previous RV analysis of the system 
\citep{wittenmyer19} identified the outer planet and constrained the minimum masses and periods of the two planets.  This used data from CORALIE \citep{mayor04} comprising 70 observations between March 1999 and March 2003, and HARPS comprising 24 observations between December 2003 and December 2013.  The minimum mass of HD~65216~b was found to be 1.295$\pm$0.062\,M$_{\mathrm{J}}$ and the period was found to be 577.6$\pm$1.3\,days with an eccentricity of $0.27\pm0.02$, while HD~65216~c had a minimum mass of 2.03$\pm$0.11\,M$_{\mathrm{J}}$, period of 5370$\pm$20\,days and eccentricity of $0.17\pm0.04$.  Since that study, more RV data has been collected with HARPS, comprising 44 observations between October 2021 and December 2022, covering almost a full period of the inner planet.  This new dataset was taken after a fibre link upgrade of the HARPS spectrograph \citep{locurto15}.  We analyse this updated dataset using a Bayesian framework implemented with \texttt{numpyro} \citep{phan2019composable} using JAX \citep{jax2018github} for gradient computation and accelerated sampling with the the No-U-Turn Sampler (NUTS) \citep{Hoffman11}.\\
For our RV model, we adopt a two-planet model where the total RV is given by
\begin{equation}
    v(t) = \sum\limits^{2}_{n=1}K_{n}[\cos(\nu_{n}(t)+\omega_{n})+e_{n}\cos\omega_{n}]\space+\space \gamma_{\mathrm{inst}}
    \label{eq:rv}
\end{equation}
where $\nu_{n}(t)$ is the true anomaly of planet $n$, and $K_{n}$ is the RV semi-amplitude given by
\begin{equation}
    K_{n} = \frac{2\pi a_{n}}{P_{n}}\frac{M_{n}}{M_{\star}+M_{n}}\frac{\sin i}{\sqrt{1-e_{n}^{2}}},
    \label{eq:k}
\end{equation}
where $M_{n}$ and $M_{\star}$ are the masses of planet $n$ and the star respectively, in M$_{\odot}$, $a_{n}$ and $P_{n}$ are the planet's semi-major axis and period respectively.  In our model, we solve for the period ($P$), mass ratio ($M$/$M_{\star}$), eccentricity ($e$), argument of periastron ($\omega$) and initial phase ($\phi_{0}$) for both planets, as well as the instrument offset ($\gamma$) and the RV jitter ($\sigma$) for each dataset.\\
The eccentricity is sampled from a $\beta$ prior \citep{johnson1995beta} with parameters $a=0.867$ and $b=3.03$ given by \citet{kipping13}, which favours favour low-to-moderate eccentricities while still allowing highly eccentric solutions.  The reference time for $\phi_{0}$ is set at a Julian date (JD$_{0}$) of 2457388.5 which corresponds to the epoch time of \textit{Gaia} DR3 (2016.0) in order to simplify combining this study with astrometry in Section~\ref{sec:astro}.  The initial phase is then converted into a time of periastron by
\begin{equation}
    T_{0} = \mathrm{JD}_{0}-\frac{\phi_{0}}{2\pi}P.
\end{equation}
For convenience, we report the earliest periastron passage occurring within the RV observing window.  The priors and resultant values for orbital parameters of both planets are shown in Table~\ref{tab:rv_result}, along with the RV offset and jitter.  Note the mass ratio is sampled from a log uniform prior, between ratios of $10^{-4}$ and $10^{-1}$.  Since we keep the host mass fixed at 0.9\,M$_{\odot}$, this corresponds to a prior of $\log\mathcal{U}(0.09,90)$ for Msin$i$ in Jupiter masses, which is what we report in the table.  The semimajor axis and periastron time are not fitted parameters, but quantities derived from period and initial phase respectively.  Hence, no priors are listed for these parameters.  The offset and jitter is labelled with the subscript COR for the CORALIE dataset and H03 and H15 for the HARPS datasets before and after the fibre link upgrade respectively.
\begin{table}
\caption{Results from RV analysis.}
\renewcommand{\arraystretch}{1.0}
\begin{tabular}{l c c c c}
\label{tab:rv_result}
Parameter & Prior & b & c\\
\hline
Msin$i$[M$_{\mathrm{J}}$] & $\log\mathcal{U}(0.09,90)$ & 1.27$^{+0.07}_{-0.09}$ & 2.00$^{+0.31}_{-0.23}$\\
P[days] & $\log\mathcal{U}(10^{2},10^{4})$ & 577.9$\pm 0.5$ & 6689$^{+326}_{-259}$\\
$a$[AU] & -- & 1.31$\pm 0.00$ & 6.71$^{+0.22}_{-0.17}$\\
$e$ & $\beta(0.867, 3.03)$ & 0.25$^{+0.04}_{-0.03}$ & 0.58$^{+0.06}_{-0.05}$\\
$\omega$[$\degree$]  & $\mathcal{U}(0,360)$ & 225$^{+11}_{-13}$ & 116$^{+14}_{-16}$\\
$\phi_{0}$[$\degree$]  & $\mathcal{U}(0,360)$ & 85$^{+12}_{-10}$ & 180$^{+9}_{8}$\\
T$_{0}$[JD] & -- & 2451747$^{+19}_{-16}$ & 2454211$^{+279}_{-104}$\\
$\gamma_{\mathrm{COR}}$[m/s] & $\mathcal{U}(-100,100)$ & \multicolumn{2}{c}{-16.9$^{+1.7}_{-1.9}$}\\
$\gamma_{\mathrm{H03}}$[m/s] & $\mathcal{U}(-100,100)$ & \multicolumn{2}{c}{-15.4$^{+1.8}_{-1.9}$}\\
$\gamma_{\mathrm{H15}}$[m/s] & $\mathcal{U}(-100,100)$ & \multicolumn{2}{c}{-14.6$\pm 1.8$}\\
$\sigma_{\mathrm{COR}}$[m/s] & $\mathcal{U}(0,20)$ & \multicolumn{2}{c}{6.3$^{+1.5}_{-1.3}$}\\
$\sigma_{\mathrm{H03}}$[m/s] & $\mathcal{U}(0,20)$ & \multicolumn{2}{c}{6.3$^{+2.9}_{-2.7}$}\\
$\sigma_{\mathrm{H15}}$[m/s] & $\mathcal{U}(0,20)$ & \multicolumn{2}{c}{7.1$^{+2.1}_{-1.6}$}\\
\end{tabular}
\end{table}
As shown in Table~\ref{tab:rv_result}, the period of HD~65216~b agrees with the previous study, while HD~65216~c is found to have a significantly longer period with higher uncertainty. The most notable change is that HD~65216~c now has a significantly higher eccentricity, with a peak value of 0.58.  This is probably due to the sparse RV measurements between JD $\sim$2454000 and 2459000, which weakens constraints on curvature, and allows for long-period, high-eccentricity solutions.  This gap in data also allows for a wider range of solutions for RV amplitude, which could account for the larger uncertainty in mass compared to \citet{wittenmyer19}.  The fitted RV curves and the phase-folded curves for both planets are shown in Figure~\ref{fig:rv_result}.  The best fit values used for the solid curves are shown in Appendix~\ref{sec:tables} in Table~\ref{tab:rv_best}.\\
\begin{figure}
    \centering
    \subfigure[Radial velocity fit, with the offset $\gamma_{\mathrm{inst}}$ subtracted.]{\label{fig:rv_fit} \includegraphics[width=\linewidth]{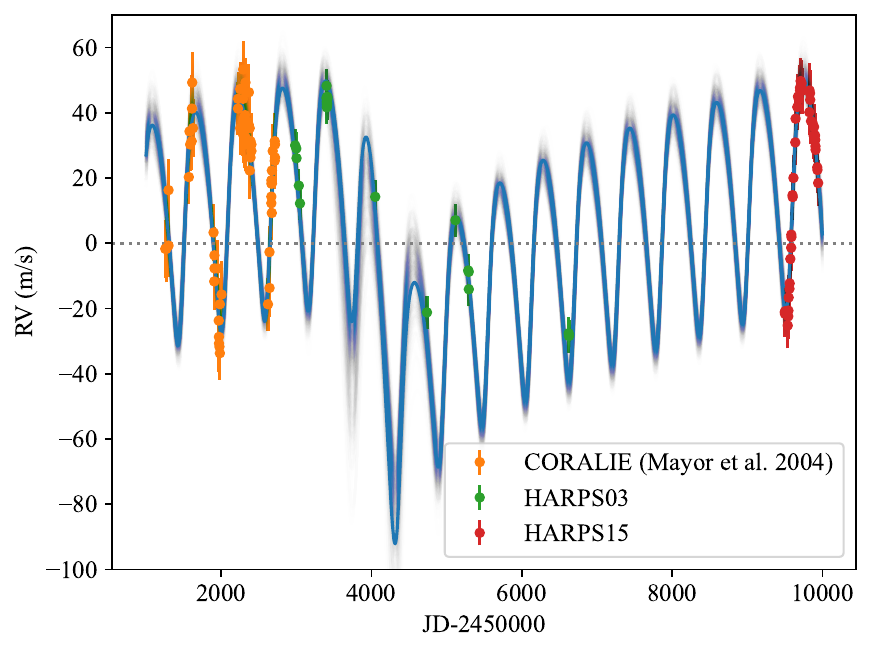}}
    \subfigure[Phase-folded curves.]{\label{fig:phase} \includegraphics[width=\linewidth]{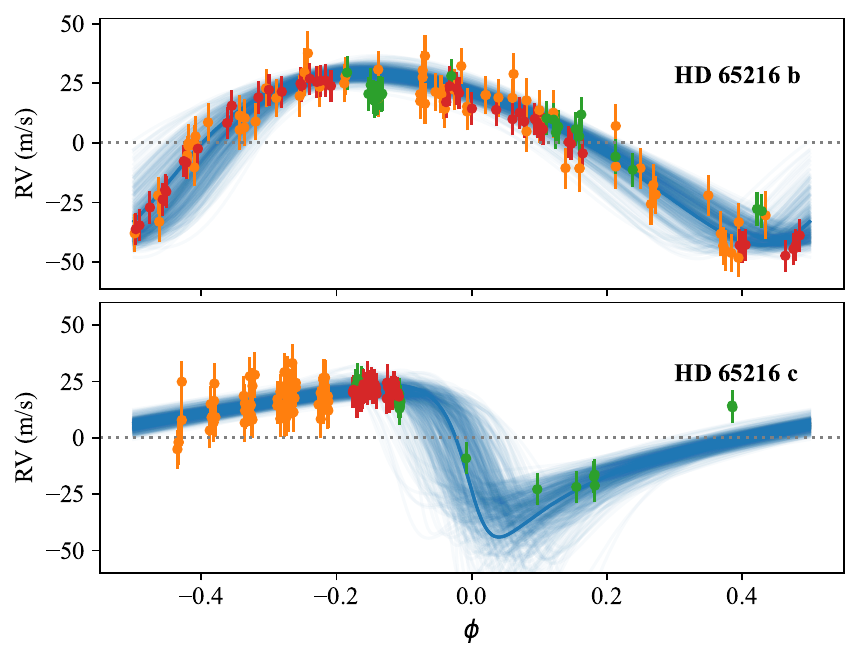}}
    \caption{Radial velocity fits and phase-folded curves for each planet.  Faint lines are from a random draw of 500 samples and dark lines are the best fit values.}
\label{fig:rv_result}
\end{figure}
Figure~\ref{fig:rv_fit} shows that, although our two-planet solution can explain the measured RV time series, there is a long ($\sim$2000\,days) gap in the data which makes precise measurement of the outer planet difficult.  The posterior for HD~65216~c is slightly asymmetric, with a tail extending towards periods exceeding 8000,days, allowing for solutions in which the outer planet has not yet completed a full orbital cycle within the RV baseline.  Additional RV monitoring over the coming decade will therefore be important for refining the orbital architecture of the system.\\
However, all our solutions for HD~65216~c's period place it within the ideal period range for a long-term astrometric study, which is detailed in the next section.
\section{Astrometric Analysis}
\label{sec:astro}
\subsection{Detection Threshold}
\begin{figure}
\centering
\includegraphics[width=1\textwidth]{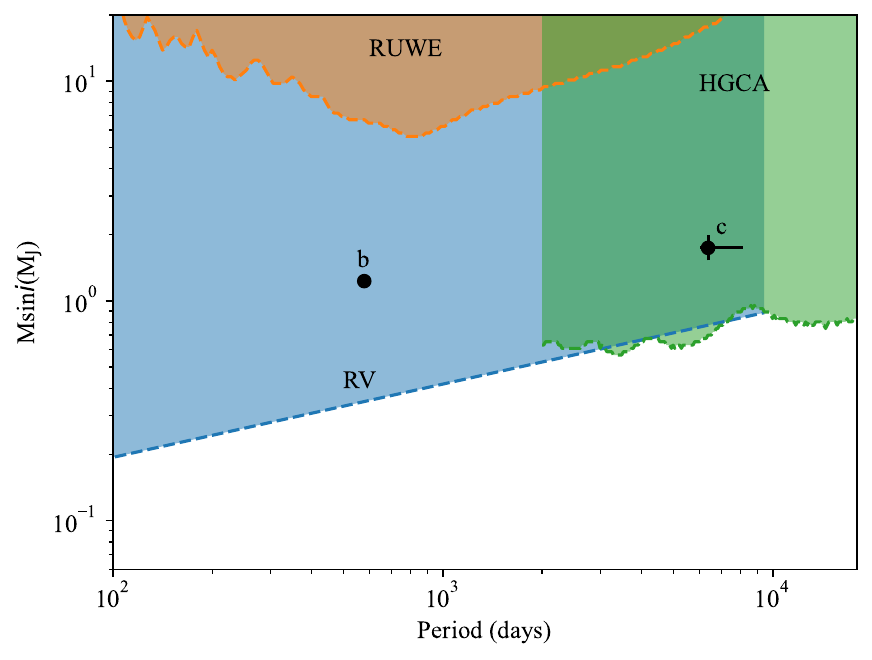}
\caption{Minimum detectable mass as a function of period for three methods. The detected planets from the radial velocity analysis are shown for comparison.}
\label{fig:threshold}
\end{figure}

The gravitational pull of a planet causes an offset of its host star's photocentre which is given by
\begin{equation}
    a^{*} = \frac{a}{d}\left(\frac{M_{p}}{M_{\star}+M_{p}}\right),
    \label{eq:offset}
\end{equation}
where $a$ is the semimajor axis of the planet and $d$ is the distance from  Earth.  For HD~65216~b and c, this offset is equal to 0.051\,mas and 0.357\,mas respectively, based on our mass and period results from RV analysis, assuming an edge-on ($i$=90\,\degree) orbit.  This implies that HD~65216~b is likely below the detection threshold for astrometry.  HD~65216~c would produce a significant offset, but its long period hinders direct detection through \textit{Gaia} DR3 alone.  This can be solved by using the long \textit{Hipparcos}-\textit{Gaia} baseline of 25\,years.\\
The detectability of planets with \textit{Gaia} DR3 alone can be quantified by the re-normalised unit weight error (RUWE) a planet's motion could induce.  This is a dimensionless quantity that represents the error in \textit{Gaia}'s fitted position, parallax and proper motion.  Reflex motion due to a companion can disrupt this fit and increase the RUWE value.  A value greater than 1.25 could be indicative of a companion  \citep{penoyre22,castro24}.  We therefore take this as our \textit{Gaia} DR3 detection threshold.  From this, we can calculate the minimum detectable companion mass as a function of period.\\
The minimum detectable mass as a function of period is shown in Figure~\ref{fig:threshold} for RV, \textit{Gaia}'s RUWE, and the \textit{Hipparcos}-\textit{Gaia} Catalogue of Accelerations (HGCA).  We define as our detection thresholds, RV semi-amplitude >10\,m/s, RUWE>1.25, and proper motion anomaly magnitude greater than the combined error of 0.064\,mas/yr.
From Figure~\ref{fig:threshold}, it can be seen that, although HD~65216~b has a period in the ideal range for detection by \textit{Gaia} DR3, its mass is below the detection threshold, if we assume an edge-on orbit with mass equal to the minimum mass.  This is also reflected in the RUWE value of 1.09, which is below the accepted companion detection threshold for DR3.  However, HD~65216~c's orbital period sits within the ideal range for detection with the \textit{Hipparcos}--\textit{Gaia} proper motion anomaly.  The sensitivity of the HGCA arises because the $\sim25$ year Hipparcos–Gaia baseline samples a substantial fraction of the outer companion’s orbit, allowing the orbital acceleration to manifest as a measurable proper motion discrepancy.  For HD~65216, the proper motion anomaly given by the HGCA \citep{brandt21} is
\begin{equation}
    \Delta\mu_{\alpha*} = 0.14\pm0.05\,\mathrm{mas/yr},
\end{equation}
\begin{equation}
    \Delta\mu_{\delta} = -0.11\pm0.04\,\mathrm{mas/yr},
\end{equation}
which is given by the long baseline proper motion between \textit{Hipparcos} and \textit{Gaia} ($\boldsymbol{\mu}_{HG}$) subtracted from the short term proper motion recorded by \textit{Gaia} ($\boldsymbol{\mu}_{G}$).  This indicates a significant acceleration which hints at the presence of a detectable long period companion.\\
\subsection{Proper Motion Modelling}
In order to calculate the planets' true masses, we rerun our \texttt{numpyro} analysis, this time including the inclination ($i$), longitude of ascending node ($\Omega$), and true proper motion as free parameters.  As input data, in addition to the RV measurements, we include the proper motion calculated by \textit{Gaia}, \textit{Hipparcos} and by the long-term change in position ($\boldsymbol{\mu}_{G}$, $\boldsymbol{\mu}_{H}$ and $\boldsymbol{\mu}_{HG}$ respectively) and associated uncertainties taken from the HGCA.  These uncertainties are used to construct a 6$\times$6 covariance matrix.\\
The long baseline proper motion is modelled by calculating the offset of the star's photocentre from the centre of mass at \textit{Hipparcos} and \textit{Gaia}'s epochs times, due to the orbital motion of both planets.  We model the star's position due to planet $n$ with:
\begin{equation}
    r_{n}(t) = a^{*}_{n}\left[\frac{1-e_{n}^{2}}{1+e_{n}\cos\nu_{n}(t)}\right],
\end{equation}
where $a^{*}_{n}$ is the average angular offset of the photocentre due to planet $n$, defined by Equation~\ref{eq:offset}.  Other parameters are defined as in Equations~\ref{eq:rv} and~\ref{eq:k}.  The offset in the R.A. ($\alpha$) and Dec. ($\delta$) directions are given by:
\begin{equation}
    x_{\alpha} = \sum\limits_{n=1}^{2}r_{n}[\sin\Omega\cos(\nu_{n}+\omega_{n})+\cos\Omega\sin(\nu_{n}+\omega_{n})\cos i],
\end{equation}
\begin{equation}
    x_{\delta} = \sum\limits_{n=1}^{2}r_{n}[\cos\Omega\cos(\nu_{n}+\omega_{n})-\sin\Omega\sin(\nu_{n}+\omega_{n})\cos i].
\end{equation}
The long baseline proper motion is then given by:
\begin{equation}
    \boldsymbol{\mu}_{HG} = \frac{\boldsymbol{x}_{G}-\boldsymbol{x}_{H}}{t_{G}-t_{H}}+\boldsymbol{\mu}_{0},
\end{equation}
where $\boldsymbol{\mu}_{0}$ is the proper motion of the barycentre (the `true' proper motion) and $\boldsymbol{x}_{G}$ and $\boldsymbol{x}_{H}$ are the astrometric offsets of the photocentre at times $t_{G}$ and $t_{H}$ (2016.0 and 1991.25 respectively.)\\
The short term proper motion from \textit{Gaia} is modelled using a simplified version of the Astrometric Global Iterative Solution (AGIS) \citep{lindegren12} in which the photocentre's offsets in \textit{Gaia}'s scanning direction are modelled in order to then model the observed proper motion and parallax.  We define the photocentric offset in the scanning direction due to planetary motion as
\begin{equation}
    x_{AL,\mathrm{planets}} = x_{\alpha}\sin\psi+x_{\delta}\cos\psi,
\end{equation}
where $\psi$ is the scanning angle.  The total offset due to parallax, proper motion and planetary orbits is then given by
\begin{equation}
    x_{AL} = \boldsymbol{A}\begin{bmatrix}
\Delta\alpha* \\
\Delta\delta \\
\varpi \\
\mu_{0\alpha*} \\
\mu_{0\delta}
\end{bmatrix}+x_{AL,\mathrm{planets}},
\label{eq:xAL}
\end{equation}
where $\Delta\alpha*$ and $\Delta\delta$ are constant offsets in R.A. and Dec. respectively (assumed to be 0), $\varpi$ is the parallax, and $\mu_{0\alpha*}$ and $\mu_{0\delta}$ are the proper motions of the barycentre.The matrix $\boldsymbol{A}$ is given by
\begin{equation}
    \boldsymbol{A} = \begin{bmatrix}
\sin\psi & \cos\psi  & f_{\varpi} & (t-t_G)\sin\psi & (t-t_G)\cos\psi
\end{bmatrix},
\end{equation}
where $\psi$ is the scanning angle and $f_{\varpi}$ is the `parallax factor,' which is the projection of the barycentric parallax displacement vector onto the Gaia along-scan direction \citep{lindegren12}.  The expected observations are then calculated using the inverse of the first part of Equation~\ref{eq:xAL}:
\begin{equation}
    \begin{bmatrix}
\Delta\alpha* \\
\Delta\delta \\
\varpi \\
\mu_{\alpha*} \\
\mu_{\delta}
\end{bmatrix} = (\boldsymbol{A}^{T}\boldsymbol{A})^{-1}\boldsymbol{A}^{T}x_{AL}
\label{eq:params}
\end{equation}
For simplicity, we do not solve for the true parallax, and keep this fixed at the value reported by \textit{Gaia}.  This is because a planetary mass companion is unlikely to have a significant effect on the observed parallax.  The \textit{Hipparcos} proper motion ($\boldsymbol{\mu}_{H}$) is modelled the same way, using the \textit{Hipparcos} scanning angles and observation times.\\
Although HD~65216 is a triple system, we do not consider the effect of the two M-dwarf companions, as their separation of 253\,AU from the primary \citep{mugrauer07}, with a period of $\sim$4000\,years, makes the reflex motion negligible over the \textit{Hipparcos}-\textit{Gaia} baseline.
\subsection{Results From Joint RV-Astrometry Analysis}
\begin{figure}
\centering
\includegraphics[width=0.9\textwidth]{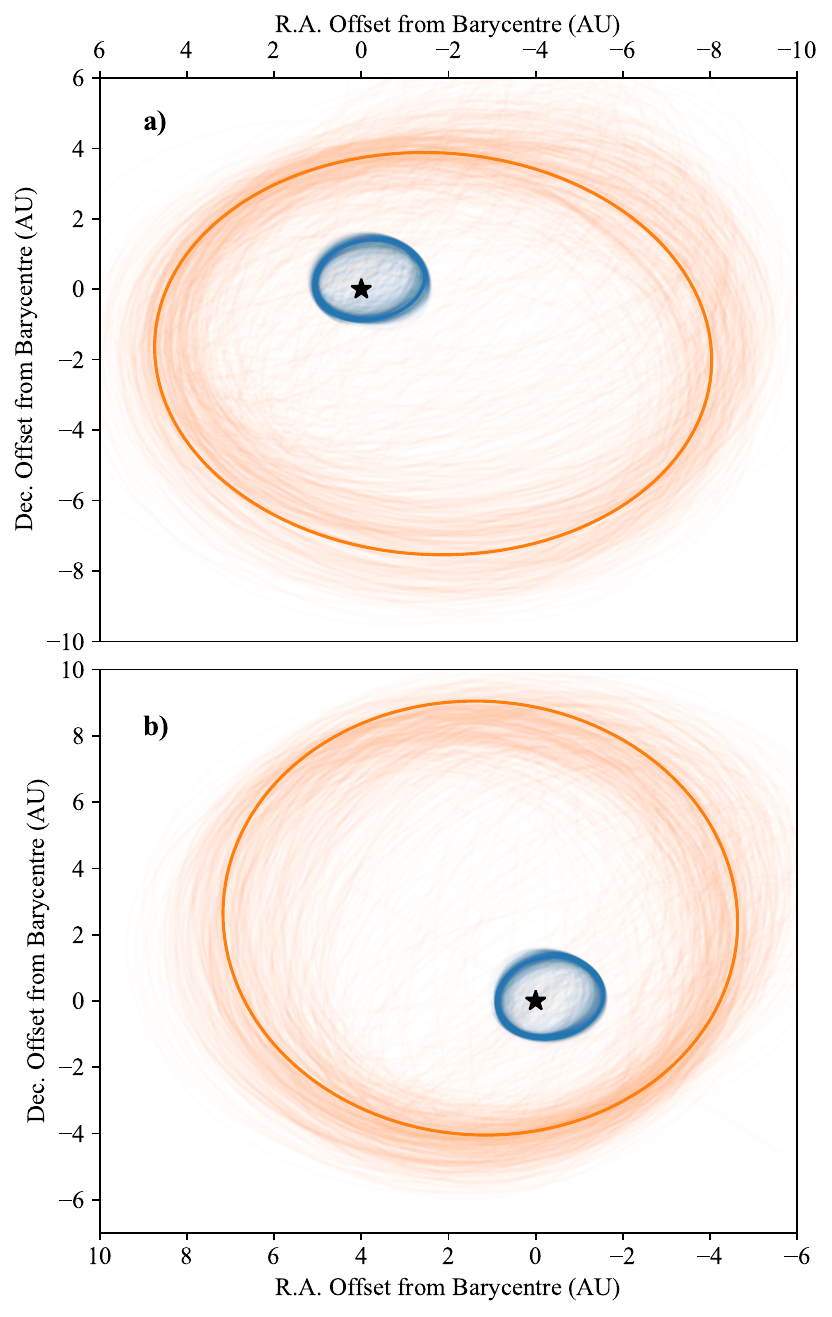}
\caption{Planet orbit solutions for anti-clockwise (\textbf{a}) and clockwise (\textbf{b}) cases.  Dark lines are best fit values and faint lines are a random draw of 500 samples.}
\label{fig:astro_vis}
\end{figure}

Given the negligible astrometric contribution expected from HD~65216~b, the current data do not meaningfully constrain mutual inclination. We therefore adopt a coplanar configuration to reduce the dimensionality of the astrometric fit and assume the same $i$ and $\Omega$ values for both planets.  We remove the RV contribution from the inner planet and conduct a combined RV-astrometry analysis on the residual signal from the outer planet.  We adopt a uniform prior in $\Omega$ spanning 0 to 2$\pi$ and a uniform prior in cos$i$ spanning -1 to 1.  We use uniform priors for the true proper motion of the system, spanning a space within 5\,mas/yr of the \textit{Gaia}-reported value.\\
The resultant posterior distributions of the outer planet's orbital parameters are shown in Appendix~\ref{sec:corners} in Figure~\ref{fig:hgca_corner}.\\
We find a bimodal distribution in inclination, which peaks at $\sim$30\,\degree and 157\,\degree, corresponding to anticlockwise and clockwise motion on the sky plane respectively.  The resultant true parameters of HD~65216~c, and the barycentric proper motion are summarised in Table~\ref{tab:astro_result}.
\begin{table}
\renewcommand{\arraystretch}{1.0}
\caption{Priors and results from HGCA analysis of HD~65216~c, removing RV contribution from the inner planet.  There are two solutions for $i$ and $\Omega$: anti-clockwise ($i<90$\,\degree) and clockwise ($i>90$\,\degree).  Both are listed, separated by commas.}
\begin{tabular}{l c c}
\label{tab:astro_result}
Parameter & Prior & Value\\
\hline
M$_{\mathrm{c}}$[M$_{\mathrm{J}}$] & $\log\mathcal{U}(0.09,90)$ & 3.32$^{+1.40}_{-1.16}$\\
P$_{\mathrm{c}}$[days] & $\log\mathcal{U}(10^{2},10^{4})$ & 7007$^{+192}_{-161}$\\
a$_{\mathrm{c}}$[AU] & -- & 6.92$^{+0.13}_{-0.11}$\\
e$_{\mathrm{c}}$ & $\beta(0.867, 3.03)$ & 0.42$\pm 0.05$\\
$i$[$\degree$] & $\cos\mathcal{U}(-1,1)$ & 35.6$^{+24.5}_{-10.8}$, 147.1$^{+10.9}_{-24.3}$\\
$\Omega$[$\degree$] & $\mathcal{U}(0,360)$ & 107$^{+22}_{-20}$, 154$\pm 18$\\
$\omega_{\mathrm{c}}$[$\degree$] & $\mathcal{U}(0,360)$ & 120$\pm 10$\\
$\phi_{0}$[$\degree$]  & $\mathcal{U}(0,360)$ & $169^{+7}_{-8}$\\
$T_{0\mathrm{c}}$[JD] & -- & 2447086$^{+231}_{-245}$\\
$\gamma_{\mathrm{COR}}$[m/s] & $\mathcal{U}(-100,100)$  & -23.6$^{+1.4}_{-0.9}$\\
$\gamma_{\mathrm{H03}}$[m/s] & $\mathcal{U}(-100,100)$  & -15.8$^{+0.6}_{-1.1}$\\
$\gamma_{\mathrm{H15}}$[m/s] & $\mathcal{U}(-100,100)$ & -17.9$\pm 1.3$\\
$\sigma_{\mathrm{COR}}$[m/s] & $\mathcal{U}(0,20)$ & 4.0$^{+1.0}_{-1.1}$\\
$\sigma_{\mathrm{H03}}$[m/s] & $\mathcal{U}(0,20)$ & 4.5$^{+1.1}_{-0.8}$\\
$\sigma_{\mathrm{H15}}$[m/s] & $\mathcal{U}(0,20)$ & 2.1$\pm 0.3$\\
$\mu_{\alpha*}$[mas/yr] & $\mathcal{U}(-128.74,-118.74)$ & -123.88$\pm 0.03$\\
$\mu_{\delta}$[mas/yr] & $\mathcal{U}(-151.60,-141.60)$ & 146.71$^{+0.04}_{-0.03}$\\
\end{tabular}
\end{table}
We recover a true mass for HD~65216~c of 3.32$^{+1.40}_{-1.16}$\,M$_{\mathrm{J}}$, placing this companion comfortably within the planetary regime.  Assuming the planets are co-planar, we apply our best-fit minimum mass of HD~65216~b (1.25\,M$_{\mathrm{J}}$) to obtain a true mass of 2.23$^{+0.96}_{-0.77}$\,M$_{\mathrm{J}}$.  We note however, that the co-planarity assumption is not necessarily realistic and is only meant to provide a first-order approximation of the inner planet's mass.  Visualisations of the planet orbits for the anti-clockwise and clockwise cases are shown in Figure~\ref{fig:astro_vis}.
\\
\section{Prospects for DR4}
\label{sec:dr4}
\textit{Gaia's} fourth data release (DR4), expected in December 2026, will contain along-scan epoch astrometry for all sources, greatly improving our ability to detect and characterise planetary systems with astrometry.  In addition to improving the precision of measurements of parallax and proper motion, the individual observations are expected to have improved astrometric precision.  Based on its apparent G magnitude, HD~65216 has an astrometric uncertainty of 0.22\,mas in DR3 \citep{lindegren20}.  This is expected to decrease to $\sim$0.16\,mas in DR4, due to the longer observing baseline of 66\,months, up from DR3's 34\,months. The value of this uncertainty can help assess the detectability of a companion.
\subsection{Epoch Astrometry}
\label{sec:epoch}
\begin{figure}
\centering
\includegraphics[width=1\textwidth]{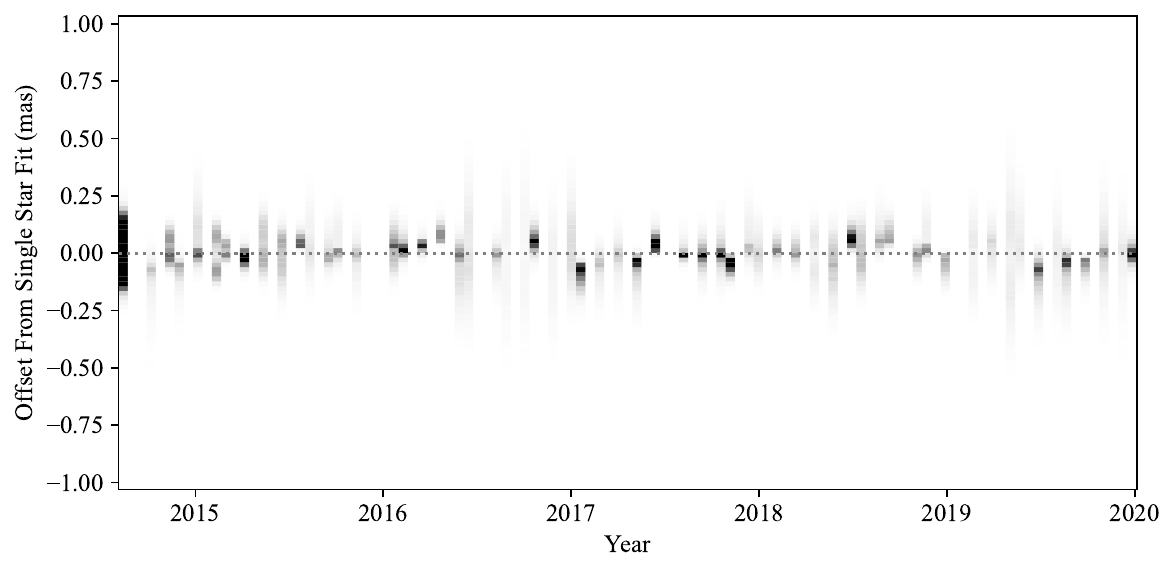}
\caption{Expected epoch astrometry from parameter distributions shown in Figure~\ref{fig:hgca_corner}.}
\label{fig:dr4_epoch}
\end{figure}

\begin{figure}
\centering
\includegraphics[width=1\textwidth]{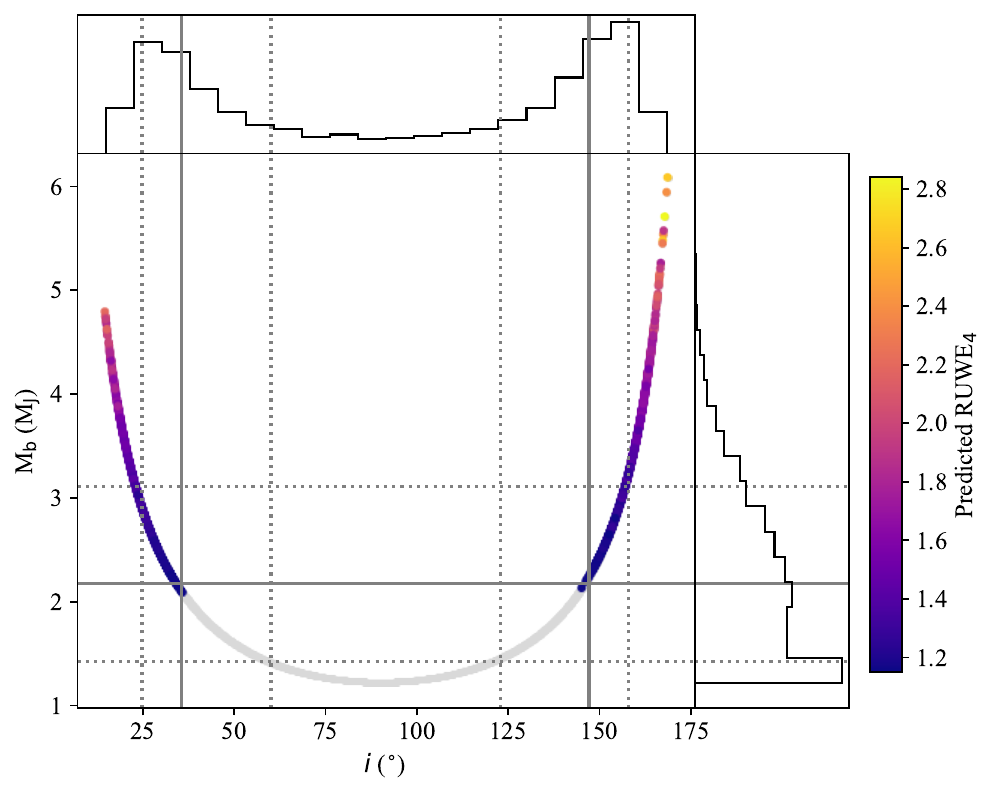}
\caption{Distribution of inclinations and mass of HD~65216~b, coloured by predicted RUWE$_{4}$.  Grey points are values of RUWE$_{4}$<1.15 and thus considered undetectable.  Posterior distributions for inclination and planet mass are shown with medians and 1-$\sigma$ values.  In the case of inclination, the medians and 1-$\sigma$ values are shown for both clockwise and anticlockwise orbits.}
\label{fig:dr4_x_samples}
\end{figure}

Using the posterior distributions of $i$ and $\Omega$, we predicted the expected along-scan offset as a function of time in DR4.  The epoch times and scanning angles are obtained from \textit{Gaia}'s Observation Forecast Tool (GOST).  For each posterior sample from our joint fit, we simulate the along-scan offset from Equation~\ref{eq:xAL} ($x_{AL}$), now including both planets and assuming co-planarity.  We then calculate the fitted parameters using Equation~\ref{eq:params}.  These are then fed back into the first half of Equation~\ref{eq:xAL} to calculate the expected offsets without planets ($x_{AL0}$).  The difference between these two offsets ($\Delta x_{AL}$) can be used as an indicator of a planet's astrometric signal.  In Figure~\ref{fig:dr4_epoch}, we show the distribution of $\Delta x_{AL}$ for each DR4 epoch.
Figure~\ref{fig:dr4_epoch} shows that, while many solutions produce small offsets within the expected along-scan error of 0.16\,mas, there are some solutions which produce significantly higher offsets.  This suggests that epoch data may be capable of further constraining inclination and, from this, true mass.  The ability of the epoch data to constrain orbital parameters can be quantified by the predicted RUWE for DR4, given by
\begin{equation}
    RUWE_{4} = \sqrt{\frac{\sum\limits_{i}^{N}(\Delta x_{i})^{2}}{\sigma^{2}(N-5)}},
\end{equation}
where $\Delta x_{i}$ are the along-scan offsets shown in Figure~\ref{fig:dr4_epoch}, $N$ is the number of epochs and $\sigma$ is the along-scan error.  In DR3, the threshold for companion detection was a RUWE of 1.25.  In DR4, this is expected to be 1.15 \citep{guerriero26}.  In Figure~\ref{fig:dr4_x_samples}, we plot the distribution of inclination and mass of HD~65216~b, coloured by the predicted RUWE$_{4}$.
As shown in Figure~\ref{fig:dr4_x_samples}, only solutions with inclination <35\,\degree or >145\,\degree produce RUWE$_{4}$ greater than the expected detectability threshold for DR4, which includes 43.5\,\% of our posterior samples.  These inclinations correspond to planet masses of 2.00\,M$_{\mathrm{J}}$ and 2.98\,M$_{\mathrm{J}}$ for HD~65216~b and c respectively.  With the release of DR4, if the astrometric offset is insignificant, these will be the upper limits of the planet masses.  If the offset is significant, we will be able to place tighter constraints on inclination and planet masses.  Additionally, it is worth noting that the inner planet HD~65216~b provides the greatest contribution to RUWE$_{4}$ as DR4 only covers a fraction of the outer planet's $\sim$19\,year orbit.  This indicates that we can use epoch astrometry to constrain the properties of the inner planet and \textit{Hipparcos}-\textit{Gaia} proper motion anomalies for the outer planet.  In this work, we assume the planets are co-planar, but a combination of these two methods will allow us to constrain the mutual inclination of the system.  It may also be possible to further constrain the properties of HD~65216~b by examining the proper motion anomaly over \textit{Gaia}'s changing baseline.
\subsection{DR3-DR4 Proper Motion Anomaly}
\begin{figure}
    \centering
    \subfigure[$i$-$\Omega$ distribution and proper motion anomaly in R.A. dimension.]{\label{fig:dpmra_inc} \includegraphics[width=\linewidth]{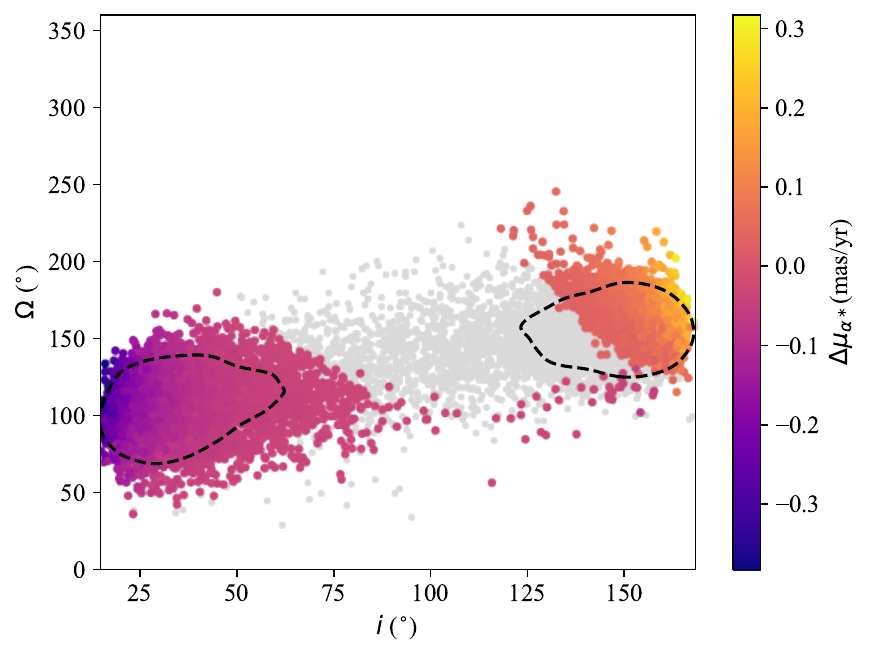}}
    \subfigure[$i$-$\Omega$ distribution and proper motion anomaly in Dec. dimension.]{\label{fig:dpmdec_inc} \includegraphics[width=\linewidth]{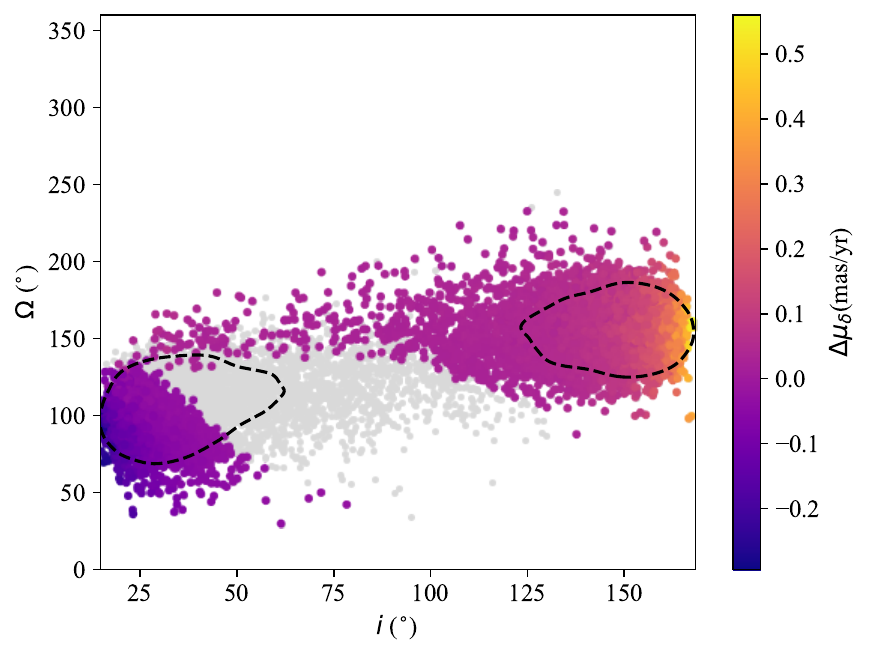}}
    \caption{Distribution of $i$ and $\Omega$, coloured by DR3--DR4 proper motion anomaly.  Grey points are solutions with anomalies below the expected uncertainties of 0.030\,mas/yr and 0.025\,mas/yr for $\Delta\mu_{\alpha*}$ and $\Delta\mu_{\delta}$ respectively.  Dashed lines show the 1-$\sigma$ level of the posterior.}
\label{fig:dr4_pma_samples}
\end{figure}

With an extra $\sim$2.7\,years of observing baseline, \textit{Gaia} DR4 will cover a larger portion of the outer planet's orbit.  As a result, the measured proper motion would be different than that measured by DR3.  This `internal' proper motion anomaly can also be used to constrain orbital parameters.\\
We simulate the expected proper motion anomaly by first simulating the epoch astrometry in DR3 and DR4 for all posterior samples, using Equation~\ref{eq:xAL}, and then calculating the proper motion fits using Equation~\ref{eq:params}.  The proper motion anomaly is then simply the difference between these fits.  In DR3, the uncertainty in $\mu_{\alpha*}$ is 0.029\,mas/yr for HD~65216 and the uncertainty in $\mu_{\delta}$ is 0.023\,mas/yr.  In DR4, these uncertainties are expected to decrease by a factor of $2\sqrt{2}$ \citep{brown25}.  We adopt these new values as the DR4 uncertainties, and calculate the  uncertainties in the proper motion anomaly $\Delta\boldsymbol{\mu}$ by combining the DR3 and DR4 uncertainties.  This is produces uncertainties of 0.030\,mas/yr and 0.025\,mas/yr for $\Delta\mu_{\alpha*}$ and $\Delta\mu_{\delta}$ respectively.\\
In Figure~\ref{fig:dr4_pma_samples}, we show the posterior distribution of inclination ($i$) and longitude of the ascending node ($\Omega$), coloured by the proper motion anomaly in both the R.A. and Dec. dimensions.
From this study, we find that 94.8\,\% of our posterior samples produce a detectable proper motion anomaly in at least one of the R.A. and Dec. dimensions.  Furthermore, Figure~\ref{fig:dr4_pma_samples} demonstrates how the sign of the DR3--DR4 proper motion is affected by the direction of the orbit.  Which suggests that, with a significant proper motion anomaly, we will go beyond the capabilities of the \textit{Hipparcos}-\textit{Gaia} anomalies and, in addition to calculating the true mass, we will distinguish between clockwise and anti-clockwise orbits.
\section{Conclusions}
\label{sec:conc}
We have presented a joint radial velocity and astrometric analysis of the HD~65216 planetary system using updated HARPS RV observations and the \textit{Hipparcos}--\textit{Gaia} Catalogue of Accelerations. Our revised RV analysis confirms the presence of the long-period outer companion HD~65216~c and favours a longer-period, more eccentric orbit than previously reported. In particular, we find evidence for a highly eccentric orbit with $e = 0.61^{+0.11}_{-0.10}$ and a period of $6455^{+273}_{-84}$\,days.\\
By combining the RV observations with the HGCA proper motion anomaly, we constrained the orbital inclination and derived true dynamical masses for both companions. The posterior distribution is bimodal, corresponding to anticlockwise and clockwise motion, with preferred inclinations near 30\,\degree and 157\,\degree respectively. The inferred true masses of HD~65216~b and c remain firmly within the planetary regime, supporting their classification as giant planets rather than brown dwarfs.\\
We further investigated the detectability of the system with upcoming \textit{Gaia} DR4 astrometry. Simulations of both epoch astrometry and the DR3--DR4 internal proper motion anomaly indicate that future \textit{Gaia} measurements should provide meaningful additional constraints on the orbital geometry of the inner planet. In particular, low-inclination (high mass) solutions are expected to produce along-scan astrometric signals comparable to or exceeding the anticipated DR4 single-epoch precision.  As the \textit{Hipparcos}-\textit{Gaia} proper motion anomaly is more sensitive to the outer planet and \textit{Gaia} epoch astrometry is more sensitive to the inner planet, these methods can be used in combination to allow constraints on the mutual inclination of the system.\\
HD~65216 therefore represents an excellent benchmark system for studying the synergy between long-baseline RV monitoring and \textit{Gaia} astrometry in the characterisation of long-period giant planets. Continued RV observations, combined with future \textit{Gaia} releases, will enable substantially improved constraints on the orbits of both planets and may ultimately permit a fully dynamical three-dimensional reconstruction of the system architecture.
\section*{Data Availability}
The data underlying this analysis are publicly available in the DACE and HGCA databases and also stored with the analysis programs on the repository \newline\hyperlink{https://github.com/awallace142857/HD65216}{https://github.com/awallace142857/HD65216}
\section*{Acknowledgements}
This work was funded by the Australian Research Council Discovery Grant DP250101273.\\
This publication makes use of The Data \& Analysis Center for Exoplanets (DACE), which is a facility based at the University of Geneva (CH) dedicated to extrasolar planets data visualisation, exchange and analysis. DACE is a platform of the Swiss National Centre of Competence in Research (NCCR) PlanetS, federating the Swiss expertise in Exoplanet research. The DACE platform is available at \href{https://dace.unige.ch}{https://dace.unige.ch}.\\
We thank the anonymous referee for their helpful comments and suggestions on how to improve this work.
\appendix
\section{Posterior Corner Plots}
\begin{figure}
    \centering
    \subfigure[Parameters of HD~65216~b.]{\label{fig:b_corner} \includegraphics[width=\linewidth]{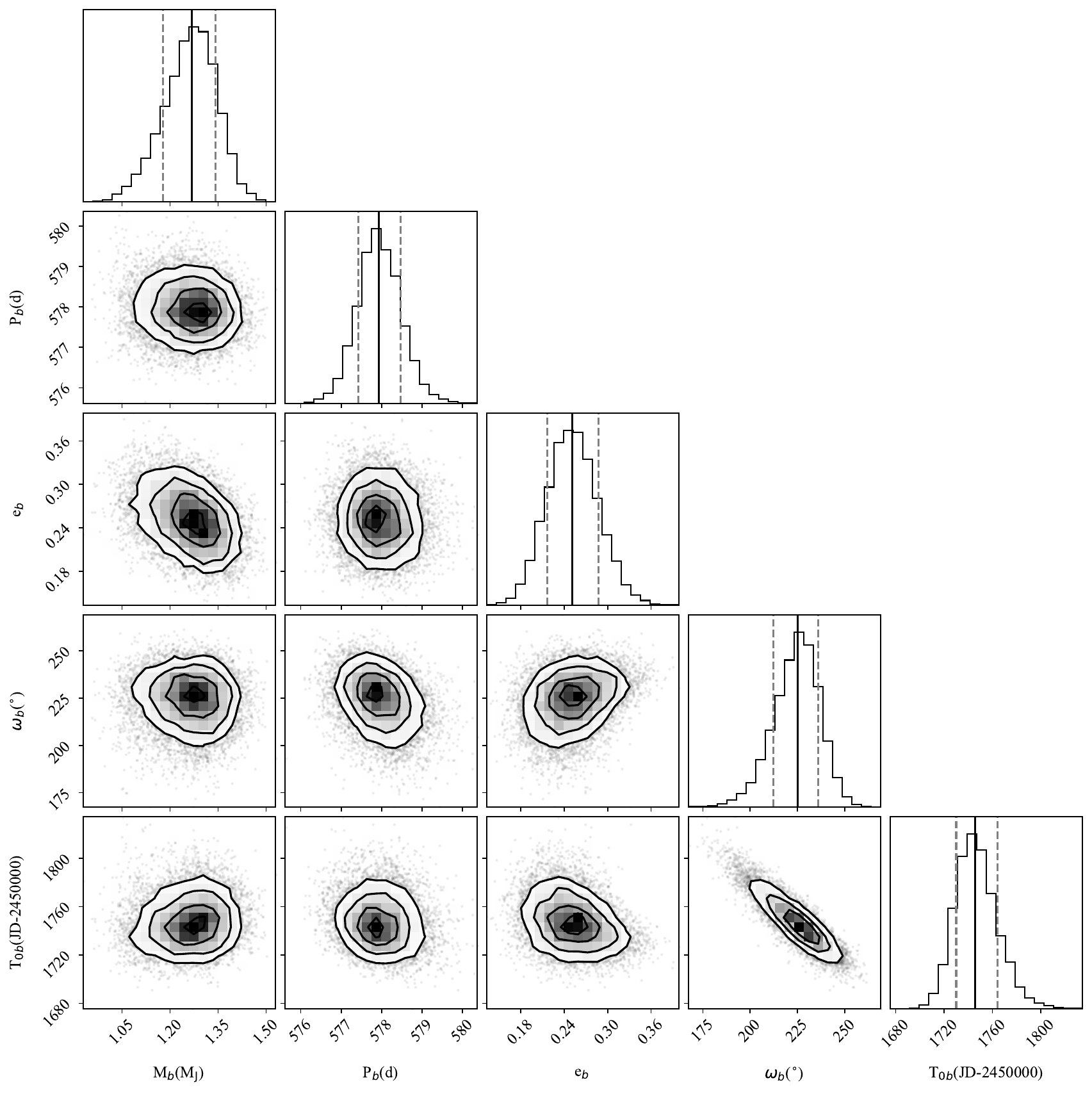}}
    \subfigure[Parameters of HD~65216~c.]{\label{fig:c_corner} \includegraphics[width=\linewidth]{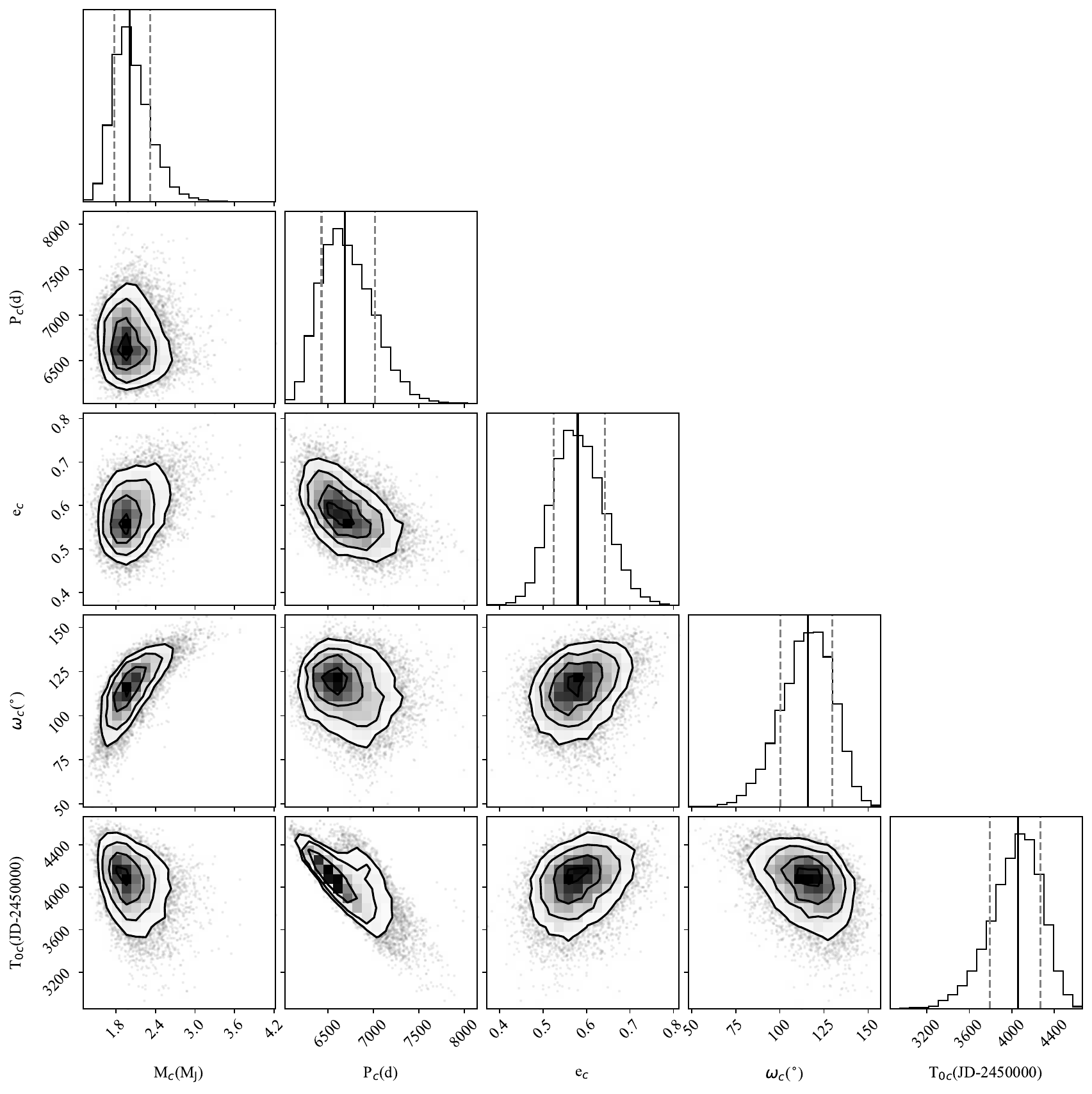}}
    \caption{Posterior distribution of planet parameters from RV analysis.  Median and 68\,\% confidence intervals are shown for each parameter.}
\label{fig:rv_corner}
\end{figure}

\begin{figure}
\centering
\includegraphics[width=1\textwidth]{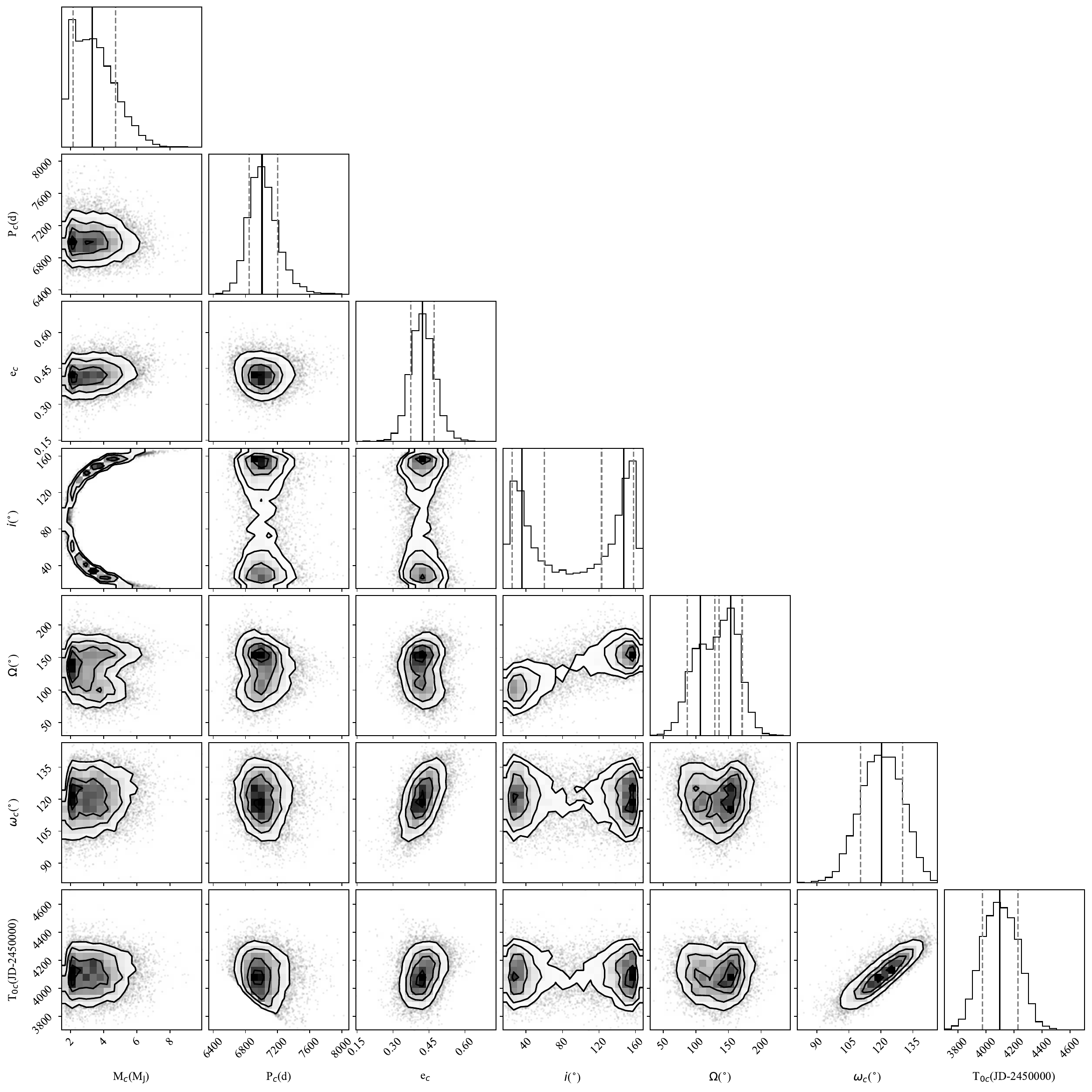}
\caption{Posterior distribution of the properties of HD~65216~c, determined from joint RV-astrometric fit.  Median and 68\,\% confidence intervals are shown for each parameter.  Due to the bimodal inclination distribution, two medians and confidence intervals are shown for $i$ and $\Omega$ distributions.}
\label{fig:hgca_corner}
\end{figure}

The posteriors of the RV fits of both planets are shown in Figure~\ref{fig:rv_corner}.
\label{sec:corners}
The posteriors of the joint RV-astrometric fit of HD~65216~c are shown in Figure~\ref{fig:hgca_corner}.
\section{Additional Tables}
Tables~\ref{tab:rv_best} and~\ref{tab:astro_result} show the best fit results of the RV and joint fits respectively.
\label{sec:tables}
\begin{table}
\caption{Maximum likelihood values from RV analysis.  These values were used for the solid curves in Figure~\ref{fig:rv_result}.}
\renewcommand{\arraystretch}{1.05}
\begin{tabular}{l c c}
\label{tab:rv_best}
Parameter & b & c\\
\hline
Msin$i$[M$_{\mathrm{J}}$] & 1.22 & 2.05\\
P[days] & 578.6 & 6637\\
$a$[AU] & 1.31 & 6.67\\
$e$ & 0.28 & 0.55\\
$\omega$[$\degree$] & 207 & 129\\
T$_{0}$[JD] & 2451761 & 2453962\\
$\gamma_{\mathrm{COR}}$[m/s] & \multicolumn{2}{c}{-16.3}\\
$\gamma_{\mathrm{H03}}$[m/s] & \multicolumn{2}{c}{-14.5}\\
$\gamma_{\mathrm{H15}}$[m/s] & \multicolumn{2}{c}{-11.4}\\
$\sigma_{\mathrm{COR}}$[m/s] & \multicolumn{2}{c}{7.6}\\
$\sigma_{\mathrm{H03}}$[m/s] & \multicolumn{2}{c}{4.8}\\
$\sigma_{\mathrm{H15}}$[m/s] & \multicolumn{2}{c}{7.5}\\
\end{tabular}
\end{table}
\begin{table}
\caption{Maximum likelihood values from RV-astrometry joint analysis of HD~65216~c.}
\renewcommand{\arraystretch}{1.05}
\begin{tabular}{l c}
\label{tab:astro_result}
Parameter & Value\\
\hline
M$_{\mathrm{c}}$[M$_{\mathrm{J}}$] & 4.92\\
P$_{\mathrm{c}}$[days] & 6931\\
e$_{\mathrm{c}}$ & 0.43\\
$i$[$\degree$] & 29.9, 156.8\\
$\Omega$[$\degree$] & 101, 153\\
$\omega_{\mathrm{c}}$[$\degree$] & 124\\
$T_{0\mathrm{c}}$[JD] & 2447150\\
$\gamma_{\mathrm{COR}}$[m/s] & -23.3\\
$\gamma_{\mathrm{H03}}$[m/s] & -15.8\\
$\gamma_{\mathrm{H15}}$[m/s] & -18.1\\
$\sigma_{\mathrm{COR}}$[m/s] & 3.7\\
$\sigma_{\mathrm{H03}}$[m/s] & 3.9\\
$\sigma_{\mathrm{H15}}$[m/s] & 2.2\\
$\mu_{\alpha*}$[mas/yr] & -123.90\\
$\mu_{\delta}$[mas/yr] & 146.75\\
\end{tabular}
\end{table}

\bibliography{references}
\end{document}